\documentclass[11pt,preprint,showpacs,preprintnumbers,amsmath,amssymb]{revtex4}

\usepackage[T1]{fontenc}
\usepackage[latin9]{inputenc}
\usepackage[a4paper]{geometry}
\usepackage{color}
\usepackage{array}
\usepackage{float}
\usepackage{multirow}
\usepackage{graphicx}
\usepackage[unicode=true,pdfusetitle,
 bookmarks=true,bookmarksnumbered=true,bookmarksopen=false,
 breaklinks=false,pdfborder={0 0 1},backref=false,colorlinks=false]
 {hyperref}
\usepackage{breakurl}

\makeatletter

\providecommand{\tabularnewline}{\\}

\@ifundefined{textcolor}{}
{%
 \definecolor{BLACK}{gray}{0}
 \definecolor{WHITE}{gray}{1}
 \definecolor{RED}{rgb}{1,0,0}
 \definecolor{GREEN}{rgb}{0,1,0}
 \definecolor{BLUE}{rgb}{0,0,1}
 \definecolor{CYAN}{cmyk}{1,0,0,0}
 \definecolor{MAGENTA}{cmyk}{0,1,0,0}
 \definecolor{YELLOW}{cmyk}{0,0,1,0}
}

\normalsize\Large

\usepackage{epsf}\usepackage{dcolumn}

\makeatother

\begin{document}

\preprint{preprint - vortex group, iitb/iitp}

\title{Effect of nominal substitution of transition metals for excess Fe
in Fe$_{1+x}$Se superconductor}

\author{Anil K Yadav$^{1}$, Santosh Kumar$^{1}$, Anup V Sanchela$^{1}$,
Ajay D Thakur$^{2}$%
\thanks{Corresponding Author \protect \\
 Email: ajay@iitp.ac.in%
}, C V Tomy$^{1}$%
\thanks{Corresponding Author \protect \\
 Email: anil@phy.iitb.ac.in%
}}

\affiliation{$^{1}$ Department of Physics, Indian Institute of Technology Bombay,
Mumbai 400076, India\\
 $^{2}$ School of Basic Sciences, Indian Institute of Technology
Patna, Patna 800013, India}
\begin{abstract}
Taking cue from the increase in the superconducting transition temperature
($T_{c}$) of Fe$_{1+x}$Se via nominal (2\,wt\%) substitution of
Cr instead of excess Fe, we have now extended our study with nominal
substitution ($\leq$5\,wt\%) with other transition metals (Ni, Co,
Fe, Mn, Cr, V and Ti) in place of excess iron. The $T_{c}$ is found
to increase (maximum $\sim11$\,K) or get suppressed depending on
the substituted transition metal. Our studies indicate that the superconducting
transition temperature depends on various parameters like the ionic
size of the transition metal, its magnetic moment as well as the amount
of hexagonal phase present as impurity.
\end{abstract}


\maketitle

\section{Introduction}

Discovery of superconductivity in LaFeAsO$_{1-x}$F$_{x}$ with a
$T_{c}\sim26$\,K in 2008 \cite{Kamihara_1} led to an outburst of
research activity towards finding new Fe-based superconductors and
increasing their $T_{c}$, which resulted in the identification of
at least six family \cite{Kamihara_1,Rottar_2,Wang_3,Hsu_4,Ogino_5,Bao_6,Stewart_7}
of Fe-based superconductors with the highest $T_{c}$ of $\sim56$\,K
reported in Gd$_{0.8}$Th$_{0.2}$FeAsO \cite{Wang2_8}. Among these
superconductors, FeSe (Fe-11) based superconductors have the lowest
$T_{c}$. Stoichiometric Fe$_{1.0}$Se$_{1.0}$ has a NiAs-type hexagonal
(space group $P6_{3}/mmc$) crystal structure and does not exhibit
superconductivity at ambient pressure. However, with a small excess
of Fe at the Fe-site (Fe$_{1+x}$Se, $x\sim0.01$), the crystal structure
gets stabilized into a tetragonal structure (space group $P4/nmm$
space group) which shows superconductivity with a $T_{c}\sim8.5$\,K
\cite{Hsu_4}. Even though the crystal structure of Fe-11 resembles
that of the other pnictides superconductors, the structure is less
complex and consists of only the alternate Fe--Se planes with no spacer
layers in between, which makes them ideal materials to investigate
the superconducting properties in Fe-based high $T_{c}$ superconductors
in general. Application of an external pressure is found to increase
the $T_{c}$ of the Fe-11 compounds \cite{Mizuguapl_9,Margadona_10,Medvedev_11}.
Even for the non-superconducting stoichiometic compound Fe$_{1.0}$Se$_{1.0}$,
superconductivity can be achieved with a $T_{c}$ as high as 27\,K
with an applied pressure of $P=1.48$\,GPa \cite{Mizuguapl_9}. The
$T_{c}$ of the non-stoichiometric tetragonal compound Fe$_{1.01}$Se
increases initially with the hydrostatic pressure, attaining a maximum
$T_{c}$ of $\sim37$\,K for $P\sim7$\,GPa and then decreases down
to $\sim6$\,K at 14\,GPa \cite{Margadona_10}. In all the pressure
effect studies on $T_{c}$, it is found that the non-superconducting
hexagonal phase increases along with the increase in $T_{c}$. As
the crystal structure gets completely transformed into the hexagonal
phase, superconductivity gets fully suppressed \cite{Medvedev_11}
implying that there is an optimum ratio between the two phases for
the maximum $T_{c}$. The alternate way to increase $T_{c}$ of the
Fe-11 compounds is the chemical pressure, achieved by the doping of
other chemical elements at the Fe or Se site. $T_{c}$ is found to
increase to a maximum of $\sim15$\,K at ambient pressure by substituting
50\% Te at the Se site \cite{Fang_12,Yeh_13,Sales_14,Taen_15}, whereas
the S substitution at the Se site is found to increase $T_{c}$, only
to a maximum of $\sim$10.5\,K \cite{Mizuguchi_16}. Wu \emph{et
al}., \cite{Wu_17} have studied the effect of substitution (more
than 10 \%) at the Fe site by non transition metals (Al, Ga, In, Sm,
Ba) and transition metals (Ti, V, Cr, Mn, Co, Ni and Cu) in the Fe$_{1+x}$Se
compound, but could not observe any enhancement in $T_{c}$. The only
other family of compounds which showed a maximum $T_{c}$ of $\sim$32\,K
amongst the FeSe-based compounds is the $A$$_{x}$Fe$_{2-y}$Se$_{2}$
($A$ = K, Cs, Rb) family of compounds, but with a different \cite{guo18,Wang19,Kazakov20}
crystal structure (ThCr$_{2}$Si$_{2}$ structure, space group $I4/mmm$).
We have earlier reported an increase in $T_{c}$ upto 11\,K by substituting
excess Cr (2\%) at the Fe site, instead of excess Fe \cite{Anilprb_21,Anilsolid_22}.
This motivated us to investigate the effect of nominal substitution
of other transition metal (TM) elements at the Fe site in the Fe$_{1+x}$Se
compound. The substitution of TMs was started with $x=0.1$ and then
subsequently increased to search for the optimal stoichiometry for
the maximum $T_{c}$ and diamagnetic shielding fraction in each substitution.
In this paper, we present the physical properties (structural, magnetization,
electrical transport and thermal transport) only for the optimally
doped Fe\emph{$T_{x}$}Se, ($T$ = Fe, Mn, Cr, V, Ti)) compounds,
which show the maximum $T_{c}$. We could not observe a clear evidence
for superconductivity when Ni and Co was substituted in place of Fe.
Our studies indicate that the superconducting transition temperature
depends on various parameters like the ionic size of the transition
metal, its magnetic moment as well as the amount of hexagonal phase
present as impurity.

\section{Experimental Techniques}

Polycrystalline samples of Fe$T$$_{x}$Se ($T$ = Ni, Co, Fe, Mn,
Cr, V and Ti) were prepared via conventional solid state reaction
method. The starting materials of high purity element powders were
taken in stoichiometric ratio and mixed in an agate mortar. The mixture
was then heated in an evacuated quartz tube at 20\,$^{\circ}$C/hour
and held at 1050\,$^{\circ}$C for 24 hours. After that, the samples
were cooled slowly down to 360\,$^{\circ}$C, held at that temperature
for 24 hours and then quenched from this temperature in liquid nitrogen
(LN2). The quenching process was adopted to minimize the formation
of $\alpha$-FeSe hexagonal phase, since the tetragonal $\beta$-phase
is known to exist above $\sim300$\,$^{\circ}$C \cite{Okamoto_23,McQueen_24}.
For studying the effect of quenching, a few samples ($T$ = Fe, Mn,
Ti and V) were also synthesized by cooling to room temperature in
the final stage instead of quenching from 360\,$^{\circ}$C. Powder
X-Ray diffraction (Cu-$K_{\alpha}$) patterns were obtained using
a Panalytical X'Pert Pro ($\theta$--$2\theta$ scans) for the structural
analysis and the phase purity determination. For the chemical identification
and the stoichiometry analysis, the energy dispersive X-ray spectroscopy
(EDAX) was performed. DC magnetization and ac susceptibility ($H_{{\rm ac}}=3.5$\,Oe
and frequency $f=211$\,Hz) measurements were carried out in a Superconducting
Quantum Interference Device - Vibrating Sample Magnetometer (SVSM)
(Quantum Design, USA). Electrical resistance using four probe method
was measured using the resistivity option of the Physical Property
Measurement System (PPMS), Quantum Design Inc., USA. Thermal conductivity
and thermopower measurements were performed via the thermal relaxation
method in the TTO option of the PPMS.

\section{Results }

\subsection{Structure analysis}

\begin{figure}[H]
\begin{centering}
\includegraphics[scale=0.45]{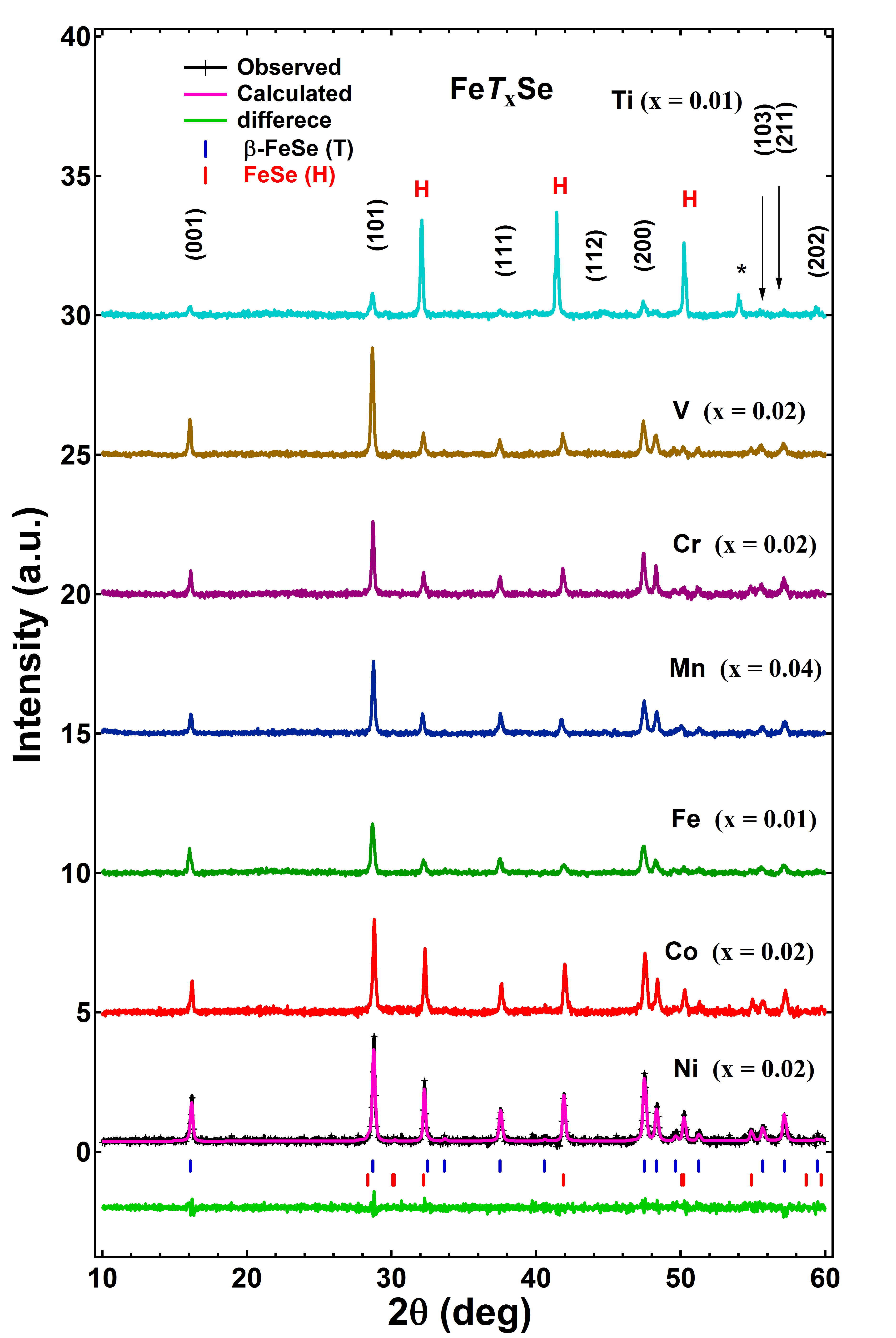}
\par\end{centering}

\caption{\label{fig:XRD_all}(Color online) Powder X-ray diffraction patterns
of all optimally doped Fe$T_{x}$Se ($T$ = Ni, Co, Fe, Mn, Cr, V
and Ti) samples. The XRD peaks which match with the tetragonal structure
are indexed with ($h,k,l$) values and the peaks that match with the
hexagonal structure are indicated with the symbol H. Asterisk denotes
other impurity peaks. }
\end{figure}

Typical room temperature X-ray powder diffraction patterns of Fe$T$$_{x}$Se
($T$ = Ni, Co, Fe, Mn, Cr, V and Ti) samples for the optimal substitution
with LN2 quenching are shown in Fig.~\ref{fig:XRD_all}. 
 The Rietveld refinements were performed for all the samples using
the FullProf software. All the observed peaks could be indexed well
only if two phases, the tetragonal $\beta$-FeSe ($P4/nmm$ space
group) and the hexagonal $\alpha$-FeSe ($P6_{3}/mmc$ space group),
were included in the refinement. Result of one such a refinement is
also shown in Fig.~\ref{fig:XRD_all} for FeNi$_{0.02}$Se. The peaks
which match with the tetragonal phase are indexed with the corresponding
($h,k,l$) values and the peaks that match with the hexagonal phase
are marked with H in Fig.~\ref{fig:XRD_all}. However, if the peaks
could not be indexed by either of the phases, then they are marked
as impurity with an asterisk. Various parameters obtained from the
refinements are given in Table~\ref{tab:Refined-results}.\textcolor{green}{{}
}
\begin{table}[H]
\caption{\label{tab:Refined-results}Lattice parameters obtained using the
two phase ($\beta$-FeSe tetragonal phase and $\alpha$-FeSe hexagonal
phase) Rietveld refinements at 300\,K from powder XRD data. $\beta$-FeSe
belongs to the $P4/nmm$ space group with atomic positions: Fe: $2a$(3/4,1/4,0),
$T$: Ni, Co, Fe, Mn, Cr, V, Ti: $2a$(3/4,1/4,0), Se: $2c$(1/4,1/4,$z$)
and $\alpha$-FeSe belongs to the $P6_{3}/mmc$ space group with atomic
positions: Fe:$2a$(0,0,0), Se: $2c$($x$,$y$,1/4). The samples
indicated with Q are quenched in LN2 from 360$^{\circ}$C and the
samples cooled in the natural way to room temperature are marked as
RT.}
\smallskip{}

\centering{}%
\begin{tabular}{|c|c|c|c|c|c|c|c|}
\hline 
Compounds & $a$ (Å) & $c$ (Å) & $c/a$ & $V$ ($\mbox{Å}^{{\rm 3}}$) & Tet. (W\%) & Hex.(W\%)  & $T_{c}$(K)\tabularnewline
\hline 
\hline 
FeNi$_{0.02}$Se (Q) & 3.774 & 5.519 & 1.462 & 78.61 & 64.1 & 35.9 & 0.0\tabularnewline
\hline 
FeCo$_{0.02}$Se (Q) & 3.771 & 5.522 & 1.464 & 78.53 & 49.6 & 50.4 & 0.0\tabularnewline
\hline 
Fe$_{1.01}$Se (RT) & 3.768  & 5.521 & 1.465 & 78.40  & 82.8 & 17.2 & 8.2\tabularnewline
\hline 
Fe$_{1.01}$Se (Q) & 3.771 & 5.520 & 1.465 & 78.27 & 83.7 & 16.3 & 9\tabularnewline
\hline 
FeMn$_{0.04}$Se (RT) & 3.770 & 5.522 & 1.464 & 78.49 & 86.1 & 13.9 & 10\tabularnewline
\hline 
FeMn$_{0.04}$Se (Q) & 3.775 & 5.527 & 1.464 & 78.61 & 63.3 & 36.7 & 10\tabularnewline
\hline 
FeCr$_{0.02}$Se(RT)\cite{Anilprb_21} & 3.773 & 5.524 & 1.464 & 78.64 & 84.6 & 15.4 & 10.5\tabularnewline
\hline 
FeCr$_{0.02}$Se(Q) & 3.767 & 5.519 & 1.465  & 78.31 & 50.1 & 49.9  & 11.0\tabularnewline
\hline 
\hline 
FeV$_{0.02}$Se(RT) & 3.772 & 5.524 & 1.464  & 78.62  & 82  & 18 & 9.5\tabularnewline
\hline 
FeV$_{0.02}$Se(Q) & 3.772 & 5.520 & 1.463  & 78.56  & 78.5  & 21.5  & 11.2\tabularnewline
\hline 
FeV$_{0.03}$Se(Q) & 3.776 & 5.508  & 1.463  & 78.01 & 10 & 90 & 9.2\tabularnewline
\hline 
FeV$_{0.05}$Se(Q) & \multicolumn{4}{c|}{No tetragonal phase} & 0 & 100 & 0\tabularnewline
\hline 
\hline 
FeTi$_{0.01}$Se(RT) & 3.773 & 5.525 & 1.464 & 78.69 & 76.3 & 23.7  & 8\tabularnewline
\hline 
FeTi$_{0.01}$Se(Q) & 3.770 & 5.518 & 1.463  & 78.43 & 15.1 & 84.9  & 11.0\tabularnewline
\hline 
\end{tabular}
\end{table}
 The lattice parameters were found to be unaffected by the nominal
substitution of the transition metal. It is not expected that such
a small percentage of substitution by various transition metal ions
will vary the lattice parameters significantly. We have confirmed
the incorporation of the transition metals into the stoichiometry
through the EDAX spectra for all the compositions (see Table~\ref{tab:EDAX}).
The EDAX results are average value of composition that shows slight
variation in stoichiometry. However stoichiometry from Rietveld refinement
is very close to starting stoichiometry of samples. A curious observation
is the fact that the percentage of the hexagonal phase increases significantly
in some cases (Mn and Ti) due to quenching, even though the quenching
process was supposed to reduce the unwanted hexagonal phase. 
\begin{table}
\caption{\label{tab:EDAX}Composition from EDAX and Rietveld analysis of as
grown Fe$T_{x}$Se samples}
\smallskip{}

\centering{}%
\begin{tabular}{|c|>{\centering}p{30mm}|c|>{\centering}p{35mm}|>{\centering}p{35mm}|}
\hline 
S. No. & Substituted element ($T$) & Starting composition & Average composition from EDAX & Rietveld refinement stoichiometry\tabularnewline
\hline 
1 & Ni & FeNi$_{0.02}$Se & Fe$_{1.1}$Ni$_{0.02}$Se$_{0.9}$ & FeNi$_{0.02}$Se\tabularnewline
\hline 
2 & Co & FeCo$_{0.02}$Se & Fe$_{0.98}$Co$_{0.02}$Se$_{0.82}$ & FeCo$_{0.02}$Se\tabularnewline
\hline 
3 & Fe & Fe$_{1.01}$Se & Fe$_{1.02}$Se$_{0.96}$ & Fe$_{0.99}$Fe$_{0.01}$Se\tabularnewline
\hline 
4 & Mn & FeMn$_{0.04}$Se & Fe$_{1.23}$Mn$_{0.05}$Se$_{0.98}$ & Fe$_{0.997}$Mn$_{0.04}$Se\tabularnewline
\hline 
5 & Cr & FeCr$_{0.02}$Se & Fe$_{1.02}$Cr$_{0.02}$Se$_{0.9}$ & Fe$_{0.99}$Cr$_{0.02}$Se\tabularnewline
\hline 
6 & V & FeV$_{0.02}$Se & Fe$_{0.95}$V$_{0.02}$Se$_{1.05}$ & Fe$_{0.99}$V$_{0.02}$Se\tabularnewline
\hline 
7 & Ti & FeTi$_{0.02}$Se & Fe$_{0.99}$Ti$_{0.02}$Se$_{0.99}$ & Fe$_{0.997}$Ti$_{0.01}$Se$_{0.99}$\tabularnewline
\hline 
\end{tabular}
\end{table}

\subsection{Magnetization}

\begin{figure}[H]
\begin{centering}
\includegraphics[scale=0.5]{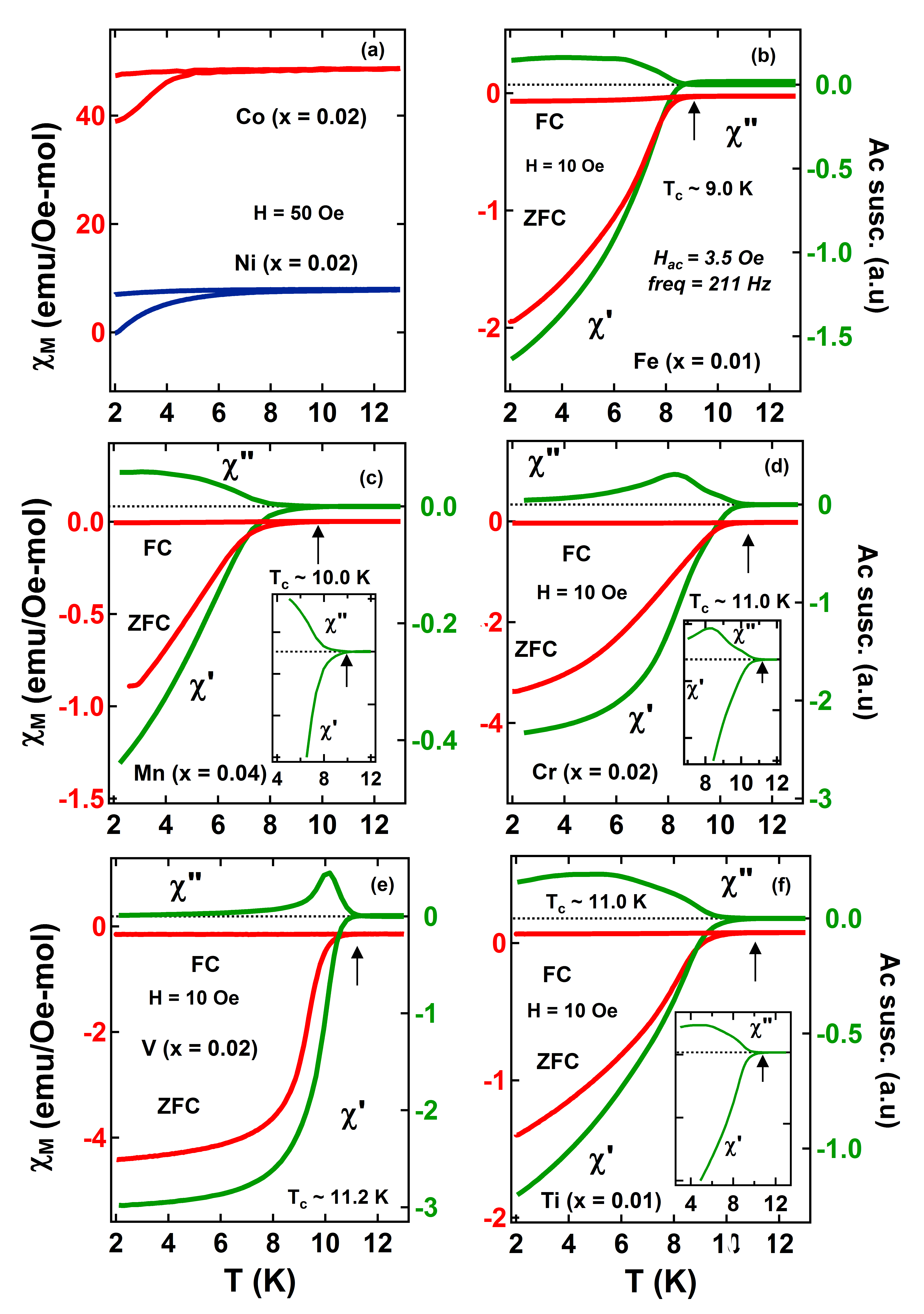}
\par\end{centering}

\caption{\label{fig:MT_all}(Color online) (a)-(f) Temperature dependence of
zero field-cooled (ZFC) and field-cooled (FC) magnetization for all
the optimally doped Fe$T_{x}$Se samples (in red color; left axis).
Temperature dependence of ac susceptibility ($\chi^{\prime}$ and
$\chi^{\prime\prime}$) for the corresponding samples is shown in
green color (right axis).\textbf{ }Insets figures show the expand
view of ac curves near $T_{c}$. }
\end{figure}

Figures\,\ref{fig:MT_all}\,(a)--(f) show the results of zero field-cooled
(ZFC) and field-cooled (FC) dc magnetic susceptibility ($\chi_{{\rm dc}}$)
for all the optimally doped transition metal-excess samples from 2\,K
to 13\,K. Superconductivity as well as transition temperature ($T_{c}$)
get suppressed in the Ni (2 wt \%) and Co (2 wt \%) excess samples
(Fig.~\ref{fig:MT_all}(a)). We see only a small dip in the magnetization
in both the Ni- and Co-excess samples. Similar type of suppression
of superconductivity was also observed by other groups with the doping
of 1\,wt\% Ni and Co in FeSe \cite{Mizuguchi_16,Wu_17,Zhang_25}.
Figure~\ref{fig:MT_all}\,(b) shows the $\chi_{{\rm dc}}(T)$ measurements
for the parent compound with excess Fe, but synthesized by the quenching
process. There is only a slight change in $T_{c}$ ($\sim9$\,K)
as compared to the $T_{c}$ of the samples or the single crystals
prepared in the usual way (slow cooling) \cite{Hsu_4,McQueen_24,Zhang(22SST)_26,Zhang(T@H SST)_27}.
Samples with the substitution of excess iron more than 1\,wt\% were
also prepared by the LN2 quenching method, but did not yield any enhancement
in $T_{c}$. The $\chi_{{\rm dc}}(T)$ data for the other transition
metal substitutions are shown in Figs.~\ref{fig:MT_all}\,(c)--(f),
where we observe an enhancement in $T_{c}$. The optimal substitution
for the highest $T_{c}$ and the shielding fraction varies for each
transition metal element; Mn -- 4\,wt\%, Cr -- 2\,wt\%, V -- 2\,wt\%
and Ti -- 1\,wt\%. The ac susceptibility measurement is usually used
as a better tool for a more precise measurement of the $T_{c}$ since
the measurement can be performed without the application of any dc
magnetic field \cite{Hein_28}. The ac susceptibility measurement
(in zero dc field) with temperature for all the substituted samples
are plotted along with the dc magnetization in Figs.~\ref{fig:MT_all}(b)--(f).
 The transition temperatures are determined from the deviation of
$\chi^{\prime}$ (in phase with applying ac signal) and $\chi^{\prime\prime}$
(out of phase) from the zero line, which are listed in Table~\ref{tab:Refined-results}.
The sharpness of the $\chi^{\prime\prime}$ peak can be taken as the
quality of the superconducting sample. Insets of Figs.~\ref{fig:MT_all}(c),(d)
and (f) show the expand portion of ac susceptibility where the bifurcation
start in $\chi^{\prime}$ and $\chi^{\prime\prime}$ to extract the
$T_{c}$.

\subsection{Electrical transport measurement }

\begin{figure}
\begin{centering}
\includegraphics[scale=0.45]{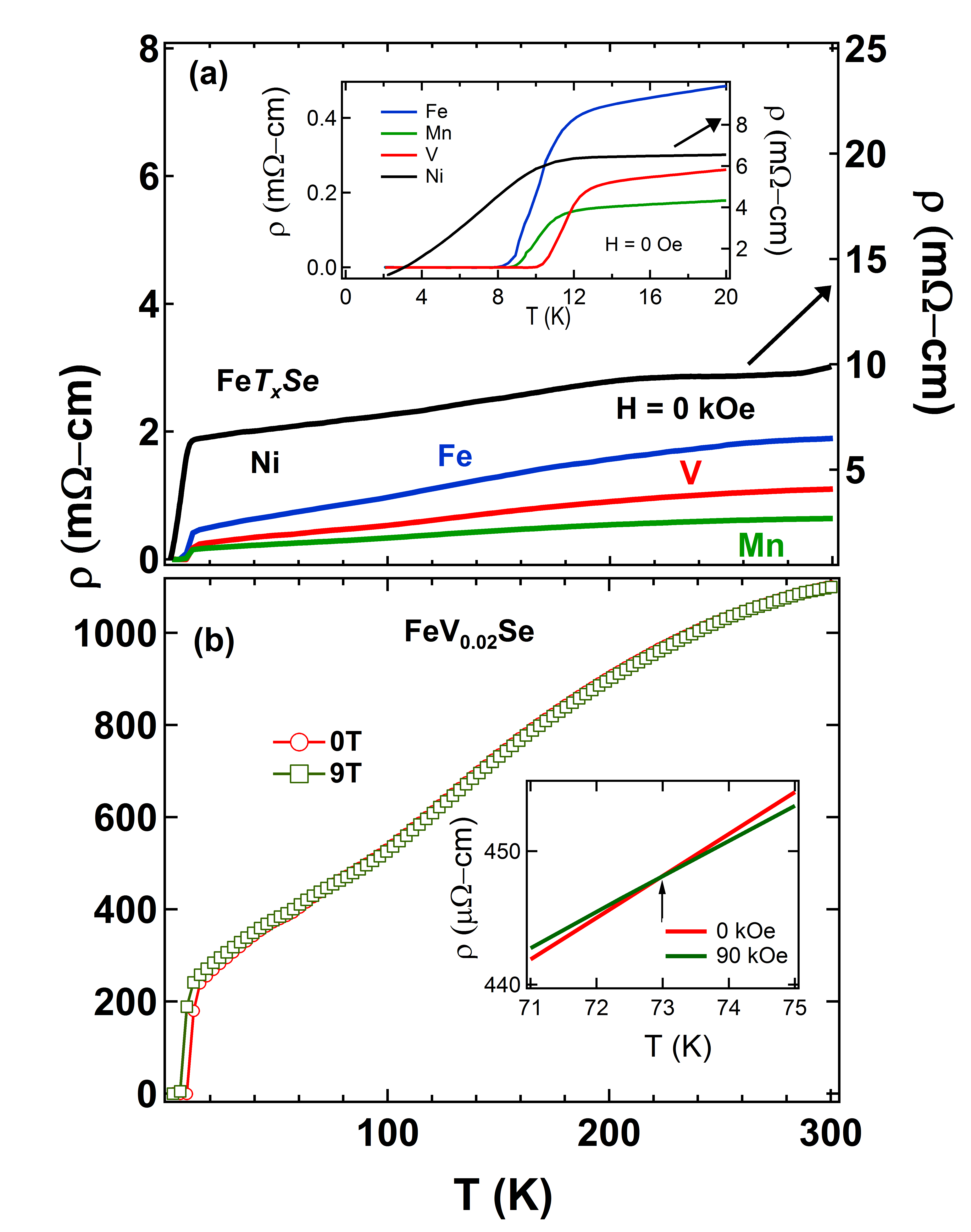}
\par\end{centering}

\caption{\label{fig:RT_V}(Color online) (a) Resistivity as function of temperature
plots in zero field for Fe$T_{x}$Se (Main panel). Inset shows the
$\rho(T)$ curves from 2\,K to 20\,K. (b) shows $\rho(T)$ curves
for FeV$_{0.02}$Se at 0\,kOe and 90\,kOe from 2\,K to 300\,K.
Inset shows the crossing of the zero and 90\,kOe $\rho(T)$ curves
(indicated by the arrow).}
\end{figure}

The main panel of Fig.~\ref{fig:RT_V}(a) shows the temperature dependence
of resistivity in zero field from 2\,K to 300\,K for all the optimally
doped Fe\emph{T}$_{x}$Se compounds, except Ti (resistance for Ti-substituted
sample could not be measured due to the brittle nature of the sample).
The resistivity curves show typical 'S' shaped curvature in the full
temperature range, which may be associated with the pseudo-gap at
the Fermi surface in these compounds at higher temperatures \cite{Song_29}.
The Fe-excess compound is found to have the highest resistivity among
all the superconducting Fe$_{1+x}$Se compounds, while the non superconducting
Ni-excess sample is found to have the largest resistivity amongst
all the compounds. The metallic characteristic of these compounds
were determined by calculating the residual resistance ratio (RRR
=$\rho_{300\,{\rm K}}/\rho_{15\,{\rm K}}$) which is given in Table~\ref{tab:Superconducting-parameters}.
These values of RRR are smaller than the values for typical metallic
conductors which suggests qualitatively that these compounds are relatively
bad conductors. The behaviour of resistance near the transition temperature
is highlighted in the inset of Fig.~\ref{fig:RT_V}(a), where the
expanded portion of the curves near the $T_{c}$ are shown between
2\,K to 20\,K. Fig.~\ref{fig:RT_V}(b) shows the typical resistivity
curves at zero field and 90\,kOe for FeV$_{0.02}$Se in entire temperature
range. The normal state resistance differs considerably from the zero
field values when the magnetic field is applied, implying a large
magneto-resistance in this compound. At about $\sim73$\,K the two
resistivity curves cross-over (see inset of Fig.~\ref{fig:RT_V}(b))
such that the resistivity which was higher for 90\,kOe now becomes
lower compared to the zero field resistivity. This crossing-over of
the curves may be associated with the structural phase transition
from the tetragonal to the orthorhombic phase observed in the low
temperature XRD measurements of Fe$_{1+x}$Se compound \cite{Hsu_4,Margadona_10,Zhang_25}.

\begin{figure}[H]
\begin{centering}
\includegraphics[scale=0.45]{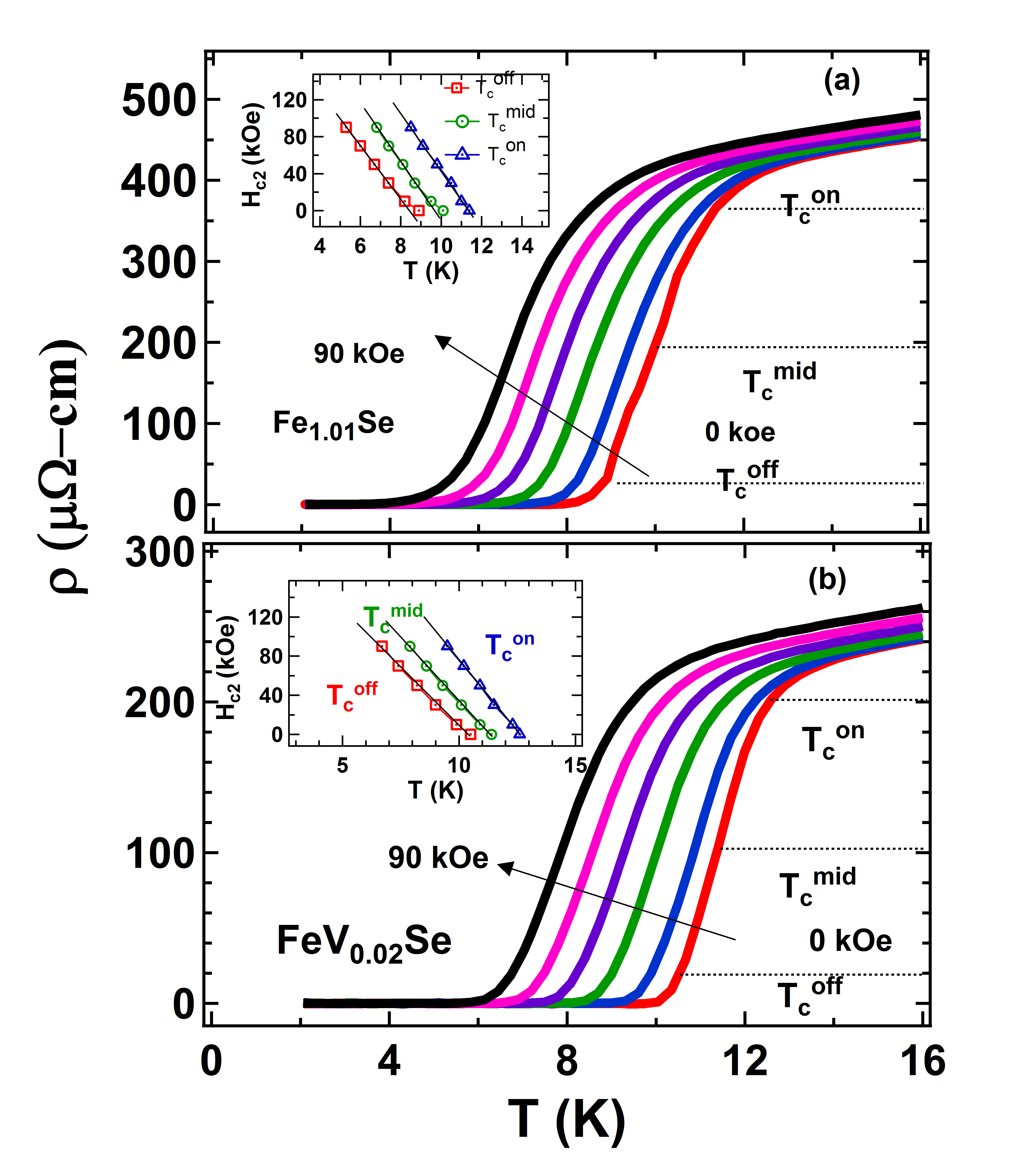}
\par\end{centering}

\caption{\label{RT_field}(Color online)(a) and (b) Temperature dependent resistivity
curves in presence of magnetic fields from 0\,kOe to 90\,kOe for
Fe$_{1.01}$Se and FeV$_{0.02}$Se samples. Insets (a) and (b) show
$H_{c2}$ vs $T$ phase diagram at three transition temperatures,
$T_{c}^{{\rm on}}$, $T_{c}^{{\rm mid}}$ and $T_{c}^{{\rm off}}$
(see text).}

\end{figure}

Fig.~\ref{RT_field}(a) and (b) show the temperature dependent $\rho(T)$
curves at various fields ranges from 0\,kOe to 90\,kOe for Fe$_{1.01}$Se
and FeV$_{0.02}$Se. We have marked three transition temperatures,
$T_{c}^{{\rm on}}$, $T_{c}^{{\rm mid}}$ and $T_{c}^{{\rm off}}$,
which are defined as 90\%, 50\% and 10\%, respectively, of the normal
state resistivity at $T=15$\,K. The upper critical field ($H_{c2}(T)$)
plots, determined at these three transition temperatures for both
samples are shown in insets of Fig.~\ref{RT_field}(a) and (b). Similar,
estimation of $H_{c2}(T)$ values for all the other compounds were
also done and the values for each superconducting sample are listed
in Table~\ref{tab:Superconducting-parameters}. The $H_{c2}(0)$
values were determined using the WHH formula \cite{Werthamer_30},
$H_{c2}(0)=-0.693T_{c}(dH_{c2}/dT){}_{T_{c}}$, where $(dH_{c2}/dT)_{T_{c}}$
is the slope at the transition temperature and are given in table
\ref{tab:Superconducting-parameters} (only for the mid transition
temperature). The Mn substituted compound has the largest $H_{c2}(0)\,(\sim236$\,kOe)
value whereas the other compounds have comparable $H_{c2}(0)\,(\sim210$\,kOe)
values. These $H_{c2}(0)$ values are comparable to the Pauli paramagnetic
limit \cite{Clogston_31} of $H_{p}=1.84T_{c}^{{\rm mid}}$, which
are also listed in Table~\ref{tab:Superconducting-parameters}. This
suggests that the spin-paramagnetic effect may be the dominant pair-breaking
mechanism in $\mbox{Fe\ensuremath{T{}_{x}}Se}$ samples as reported
for 'Fe-11' superconductors \cite{Mizuguapl_9,Lei2_32,Song_29,Chen,Khim}.
The superconducting coherence lengths ($\xi$(0)) were estimated using
the Ginzburg-Landau formula $H_{c2}$(0) = $\phi_{0}/2\pi\xi^{2}$,
which are listed in Table~\ref{tab:Superconducting-parameters}.
The $\xi(0)$ values for these substituted compounds are larger than
the Te- \cite{Lei2_32} and S-substituted compounds \cite{Mizuguchi1pl_33}.

\begin{table}[H]
\caption{\label{tab:Superconducting-parameters}Superconducting parameters
extracted from the electrical transport measurements}
\smallskip{}

\centering{}%
\begin{tabular}{|c|c|c|c|c|c|>{\centering}m{14mm}|>{\centering}m{14mm}|>{\centering}m{14mm}|c|>{\centering}m{10mm}|}
\hline 
\multirow{3}{*}{Compound} & \multicolumn{5}{c|}{$T_{c}$(K)} & \multirow{3}{14mm}{$H_{c2}^{{\rm mid}}(0)$ (kOe)} & \multirow{3}{14mm}{$\xi$(0) (nm)} & \multirow{3}{14mm}{$H_{p}^{{\rm mid}}(0)$ (kOe)} & \multirow{3}{*}{$H_{c1}$(Oe)} & \multirow{3}{10mm}{RRR}\tabularnewline
\cline{2-6} 
 & \multicolumn{3}{c|}{$T_{c}^{\rho}$} & \multicolumn{2}{c|}{$T_{c}^{\chi}$} &  &  &  &  & \tabularnewline
\cline{2-6} 
 & $T_{c}^{{\rm on}}$ & $T_{c}^{{\rm mid}}$ & $T_{c}^{{\rm off}}$ & $T_{c}^{dc}$ & $T_{c}^{ac}$ &  &  &  &  & \tabularnewline
\hline 
Fe$_{1.01}$Se & 11.4 & 10.1 & 8.9 & 8.9 & 9.0 & 210 & 3.96 & 185 & 9.5 & 4.1\tabularnewline
\hline 
FeMn$_{0.04}$Se & 11.2 & 10.1 & 9.2 & 9.8 & 10.0 & 236 & 3.73 & 185 & 10 & 3.9\tabularnewline
\hline 
FeCr$_{0.02}$Se & 13.2 & 12.0 & 11.0 & 10.9 & 11.0 & 222 & 3.98 & 220 & 24 & 8.1\tabularnewline
\hline 
FeV$_{0.02}$Se & 12.6 & 11.4 & 10.8 & 11.2 & 11.2 & 210 & 3.96 & 210 & 70 & 4.6\tabularnewline
\hline 
\end{tabular}
\end{table}

\subsection{Thermal transport properties}

\begin{figure}[H]
\noindent \begin{centering}
\includegraphics[scale=0.5]{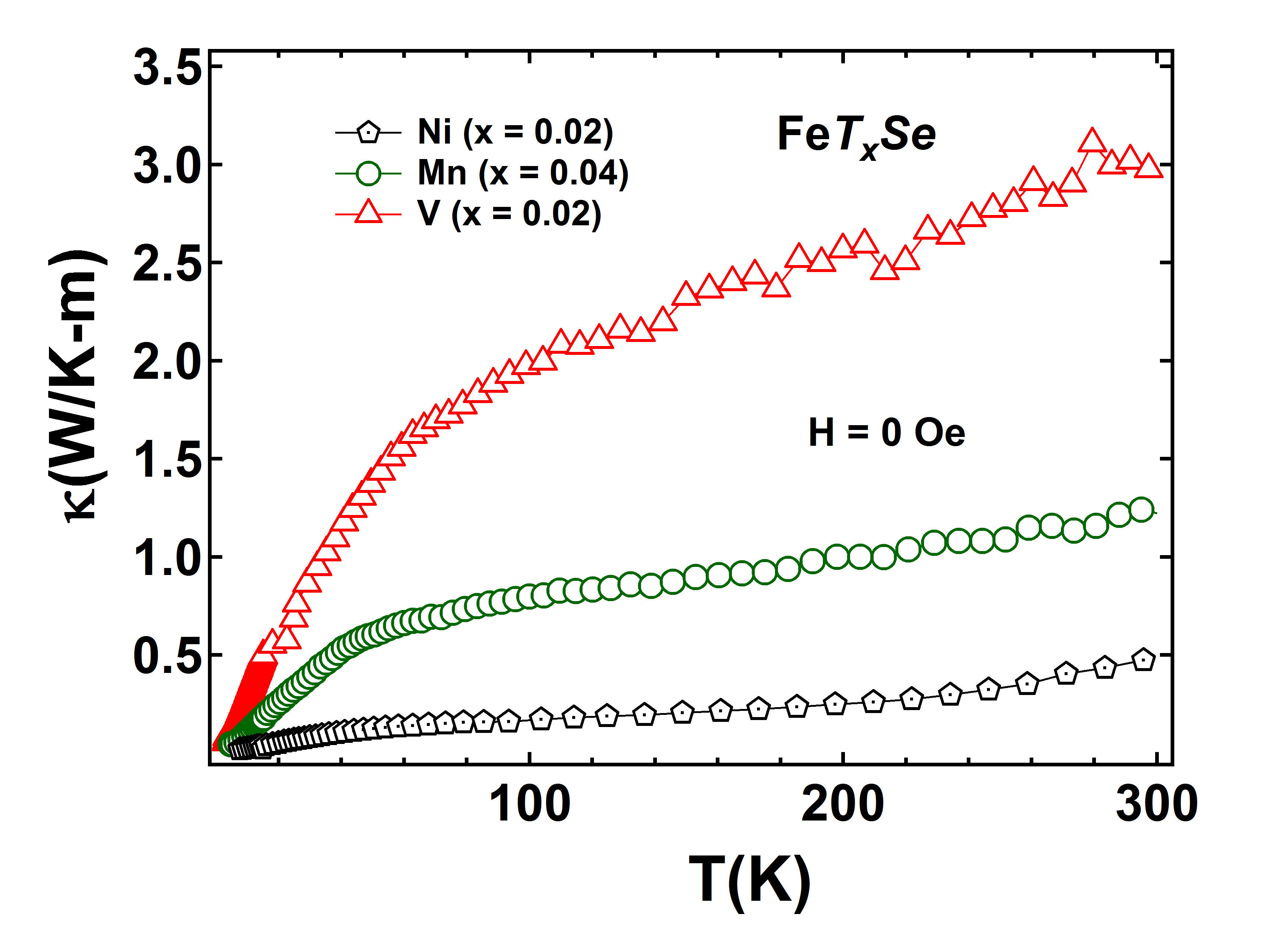}
\par\end{centering}

\caption{\label{fig:k vs T}Temperature dependence of thermal conductivity
($\kappa)$ for Fe$T$$_{x}$Se ($T$ = Ni, Mn and V) samples at zero
field. }
\end{figure}

Figure~\ref{fig:k vs T} shows the temperature dependence of thermal
conductivity ($\kappa)$ measured from 2\,K to 300\,K at zero field
for Fe\emph{T}$_{x}$Se ($T$ = Ni, Mn and V) samples. 
The $\kappa$ is the highest for FeV$_{0.02}$Se sample, which is
almost six times larger than that of the lowest $\kappa$ in FeNi$_{0.02}$Se,
however all samples have comparatively lower $\kappa$ compared to
single crystal FeCr$_{0.02}$Se \cite{Anilprb_21}. If we correlate
$\kappa$ to the number of charge carriers, then the higher value
of $\kappa$ implies larger number of charge carriers as the lattice
is the same for all these compounds. This indicates a correlation
may be exist qualitatively between the numbers of charge carriers
and in the enhancement of $T_{c}$, if we compare the $T_{c}$ (Table~\ref{tab:Refined-results})
of these compounds which varies as $T_{c}^{{\rm V}}>T_{c}^{{\rm Mn}}>T_{c}^{{\rm Fe}}>T_{c}^{{\rm Ni}}$.
On this basis charge carriers may be also have a role in the enhancement
of $T_{c}$ in Fe-11 compounds, as in the case of iron pnictide superconductors
\cite{Yang_34,Wen_35}. 

\begin{figure}[H]
\noindent \begin{centering}
\includegraphics[scale=0.5]{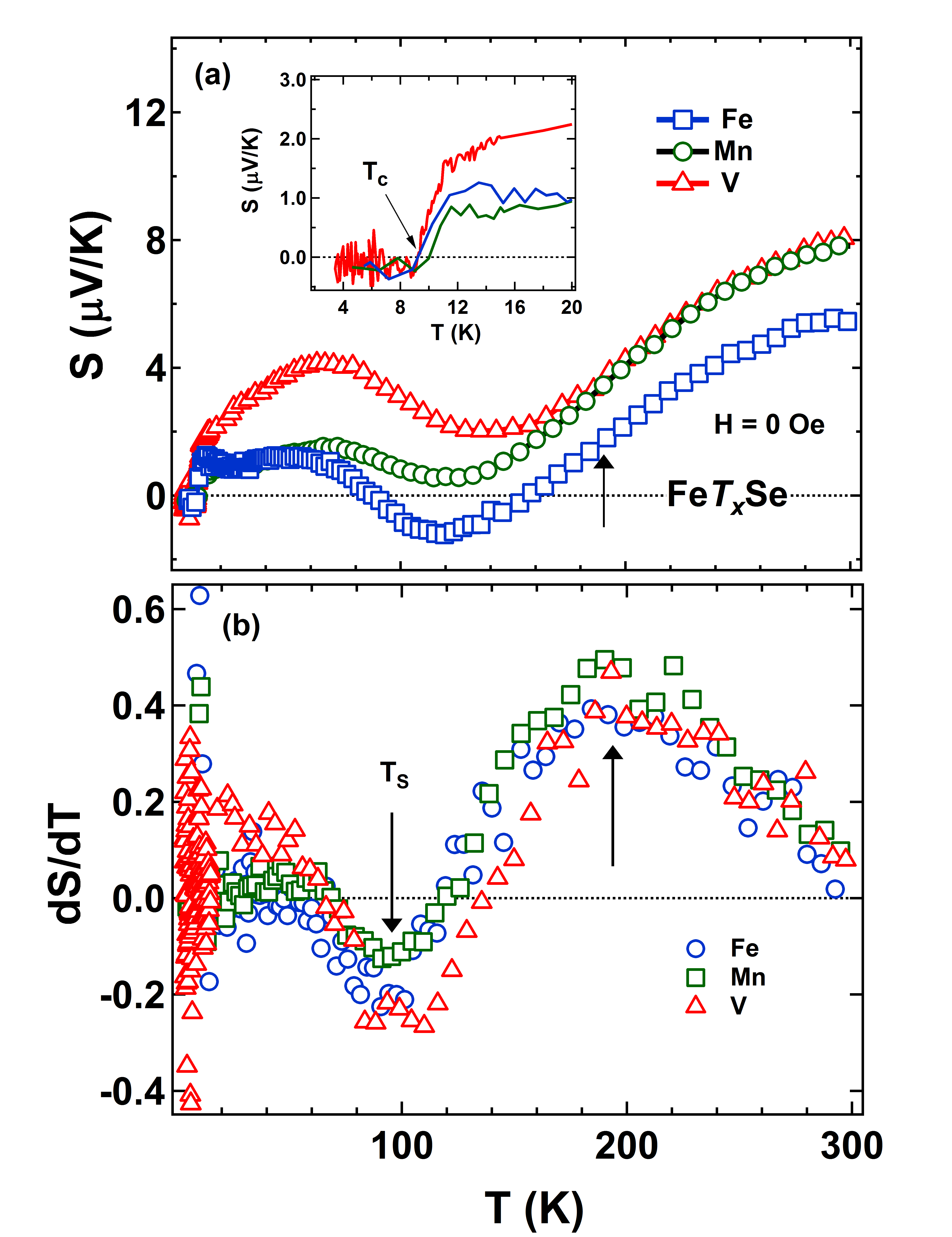}
\par\end{centering}

\caption{\label{fig:SvsT} (a)Temperature dependence of Seebeck coefficient
$(S)$ for Fe$T_{x}$Se ($T=$ Ni, Fe, Mn and V). Inset shows the
enlarged view of $S(T)$ between 2\,K to 20\,K and arrow indicates
the $T_{c}$ of compounds. (b) $dS/dT$ vs $T$ plots to highlight
the similar behaviour of transitions in the three compounds (see text).}
\end{figure}

In Fig.~\ref{fig:SvsT}\,(a), we have plotted the typical thermopower
($S$) behavior as a function of temperature for the Fe$T_{x}$Se
($T$ = Ni, Mn and V) samples. The thermopower of all the compounds
show an 'S' like curvature, typical of the Fe superconductors where
$S(T)$ decreases with decreasing temperature and goes through a minimum
between 70\,K--180\,K. It is reported that the thermopower of Fe$_{1.01}$Se
changes its sign from positive to negative at $\sim157$\,K and again
from negative to positive at $\sim$91\,K \cite{Song_29}. Similar
behaviour is seen only for the Fe$_{1.01}$Se sample in our studies.
We have observed a similar cross-over of $S$ in our earlier studies
where Cr was substituted as excess at the Fe site \cite{Anilprb_21}.
The changing of sign at $\sim93$\,K has been associated with a structural
transition in Fe$_{1.01}$Se. Since the sign of thermopower ($S$)
indicates the type of dominant charge carriers, one can argue that
the compounds, Fe$_{1.01}$Se and FeCr$_{0.02}$Se \cite{Anilprb_21},
have both type of charge carriers where as the compounds, FeV$_{0.02}$Se,
FeMn$_{0.02}$Se and FeNi$_{0.02}$Se, have only holes as major charge
carriers since the sign of $S(T)$ is positive throughout the temperature
range in these compounds. 
 Inset of Fig.~\ref{fig:SvsT}\,(a) shows an enlarged view of $S(T)$
from 2\,K to 20\,K. In the superconducting state, the Seebeck coefficient
becomes zero since the charge carriers have zero entropy as they are
involved in the Cooper pair formation. As the temperature increases,
the thermal energy overcomes the binding energy of the Cooper pairs
and the superconductor gradually enters into the normal state as shown
in the inset of Fig.~\ref{fig:SvsT}\,(a). Figure~\ref{fig:SvsT}\,(b)
shows the derivative plots of $S(T)$ to highlight the similarity
between compounds. Even though the Ni, V and Mn-substituted samples
did not show the cross-over to the negative values, their overall
behavior is the same, as evident from the derivative plots.\textbf{
}The modulation in derivative of $S(T)$ curves can be connect through
previous studies on FeSe compounds. As it is seen in theoretical \cite{Subedi,Johnston}
as well as experimental \cite{Xia} studies, Fe-superconductors are
semi-metal in nature and both types of charge carriers present at
Fermi surface. Experimentally, we have also observed the presence
of both types charge carriers in single crystals FeCr$_{0.02}$Se
\cite{Anilprb_21}. The slops change at higher temperature for these
Fe$T_{x}$Se samples occur near same temperature $(\sim$200\,K)
where types of charge carriers change in Hall measurement for single
crystal of FeCr$_{0.02}$Se \cite{Anilprb_21}. However, slop change
at lower temperature $\sim$93\,K is associated with structural transition
(tetragonal to orthorhombic) as previous reported in many references
\cite{Hsu_4,Margadona_10,McQueen_24,Pomjakushina}.

\section{Discussion}

\begin{figure}
\noindent \begin{centering}
\includegraphics[scale=0.5]{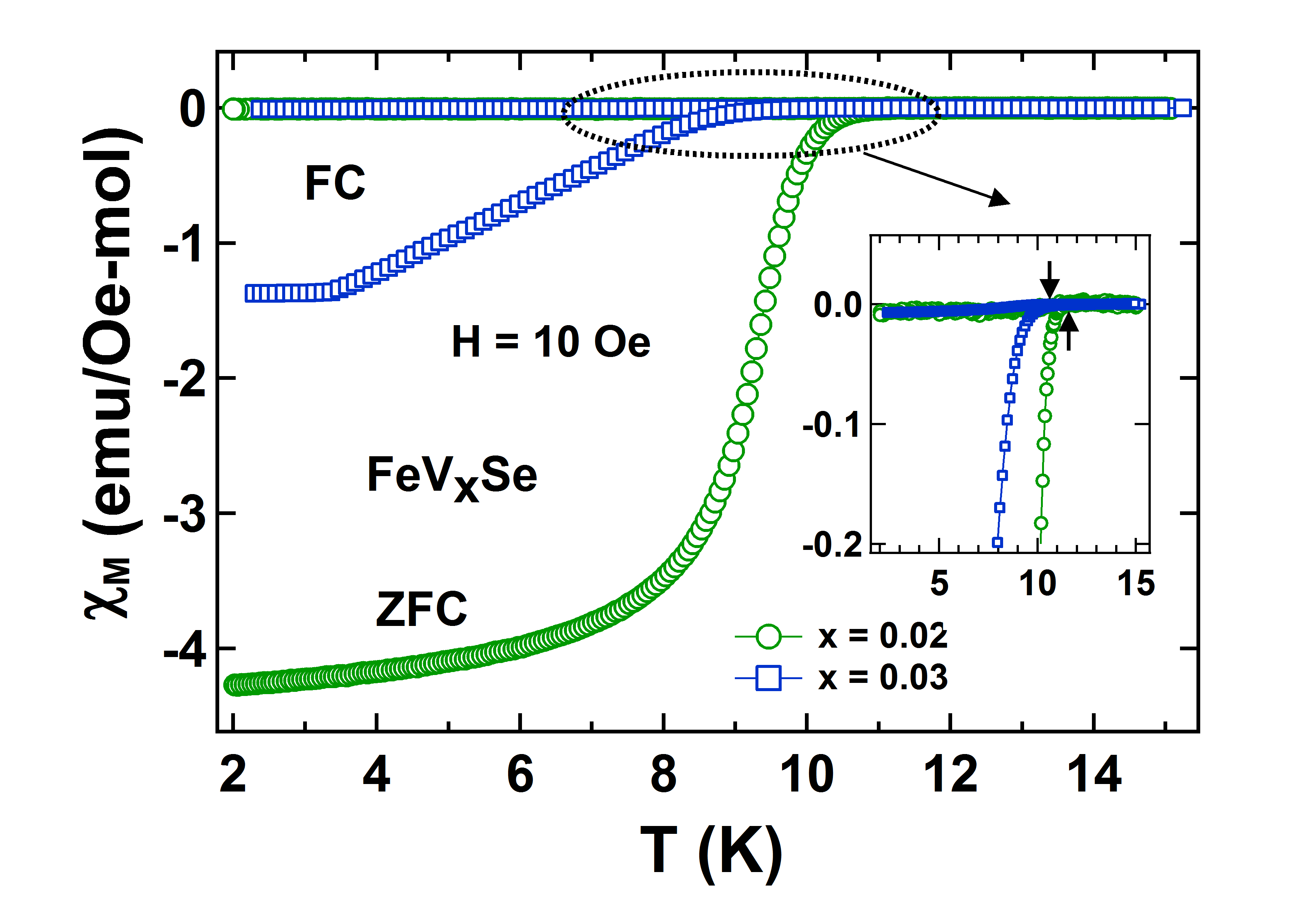}
\par\end{centering}

\caption{\label{fig:MT_V}Temperature dependence of zero field-cooled and field-cooled
plots in 10\,Oe for Fe$V_{x}$Se ($x$ = 0.02, 0.03). Inset figure
shows the enlarge view of magnetization near transition. }
\end{figure}

The quality of $\mbox{FeSe}$ samples are always a concern and a subject
of discussion since a variety of closely related phases like Fe$_{3}$O$_{4}$,
unreacted Fe, hexagonal $\mbox{FeSe}$, Fe$_{7}$Se$_{8}$, etc.,
can form a part of the prepared samples. Many groups attempted to
synthesis the pure tetragonal phase, but considerable amount of other
impurities could not be avoided \cite{Zhang(22SST)_26,Braithwaite_36}.
As per the phase diagram studied by McQueen \emph{et al.} \cite{McQueen_24},
the formation of the hexagonal (non superconducting) phase occurs
at temperatures below 200$^{\circ}\mbox{C}$. In order to minimize
this phase, we have quenched our samples from higher temperatures
(360$^{\circ}\mbox{C}$), but the percentage of hexagonal phase is
found to increase. The interesting fact is that the impurity of hexagonal
phase does not affect the $T_{c}$ of the parent compound adversely;
its presence is found to enhance the $T_{c}$ in TM-substituted compounds
(except Ni and Co). It is also found that the optimal substitution
for the maximum $T_{c}$ in each TM substituted compound is different
(Fe -- 1\,wt\%, Mn -- 4\,wt\%, Cr -- 2\,wt\%, V -- 2\,wt\%, Ti
-- 1\,wt\%). Further substitution of TM beyond the optimal value
is found to increase the hexagonal phase and decrease the tetragonal
phase. The value of wt\% substitution for the complete conversion
into the hexagonal phase is different for different substitutions
(Mn, Cr $>$ 7\,wt\%, V $\geq$ 5\,wt\%, Ti $\geq$ 2\,wt\%). It
looks as if the higher the ionic radius of the substituted TM metal,
the less wt\% of the dopant is required for the conversion into the
complete hexagonal phase. It is also found that the $T_{c}$ as well
as the diamagnetic shielding fraction decreases with increasing hexagonal
phase. As an example, we show in Fig.~\ref{fig:MT_V}, the magnetic
susceptibility data for the V-substituted samples (prepared by LN2
quenched method). The diamagnetic shielding fraction drastically gets
decreased along with a decrease in $T_{c}$ as the V-substitution
increases beyond 2\,wt\%. For the V substitution beyond 5\,wt\%
or more, the XRD patterns indicate the presence only the hexagonal
phase (see Table~\ref{tab:Refined-results}). 

\begin{figure}
\begin{centering}
\includegraphics[scale=0.55]{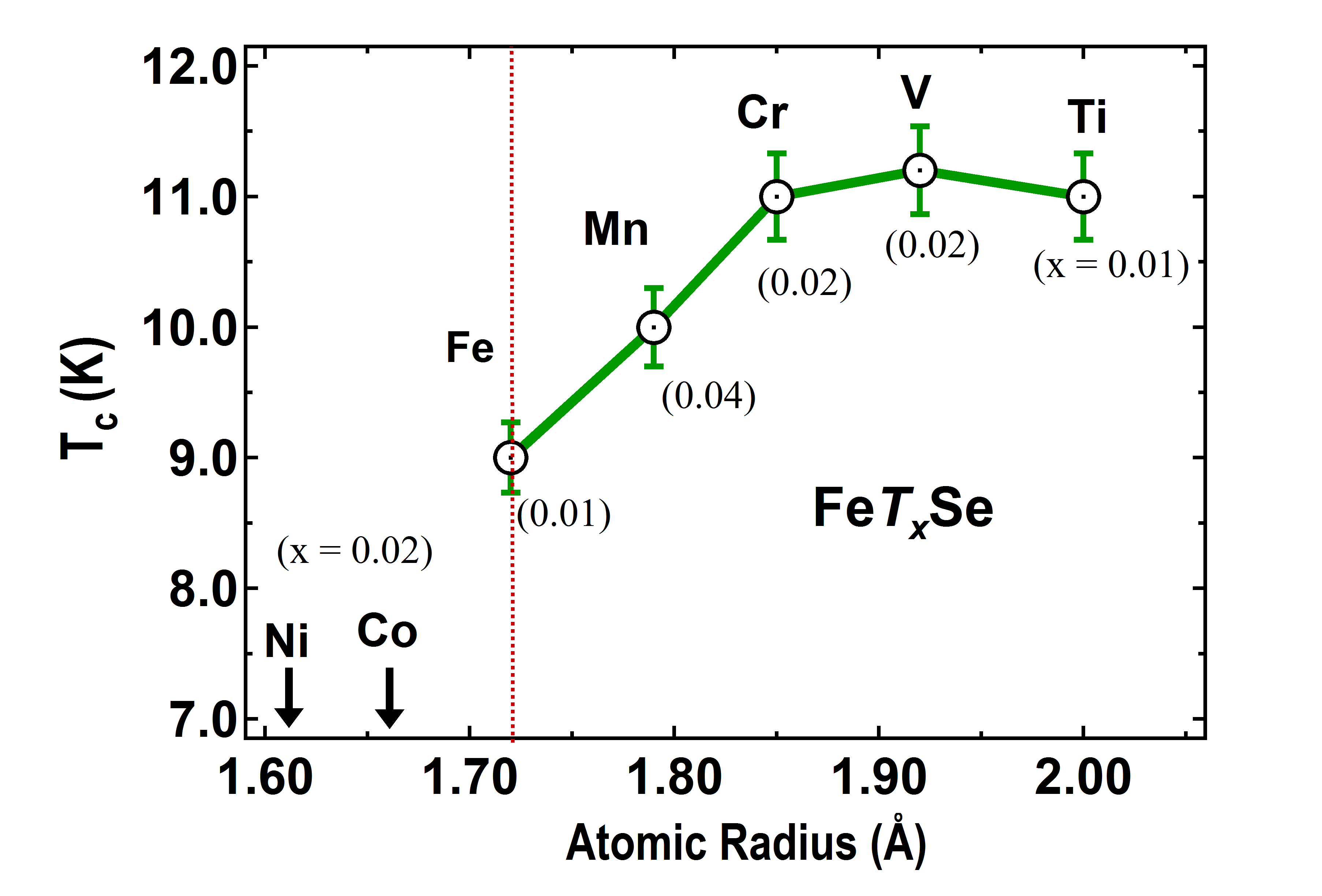}
\par\end{centering}

\caption{\label{fig:Variation-of-}Variation of $T_{c}$ with atomic radius
of the substituted transition metal ion in Fe$T_{x}$Se compounds.
The $T_{c}$ of parent superconductor Fe$_{1.01}$Se, synthesis from
quenching process, is shown by vertical dotted line.}
\end{figure}

The variation of $T_{c}$ with atomic radius of the excess substituted
transition metal in Fe$_{1+x}$Se is shown in Fig.~\ref{fig:Variation-of-}.
There is a clear indication that the $T_{c}$ of the compound gets
suppressed or destroyed when the transition metal with atomic radius
less than that of Fe is substituted as excess, where as the $T_{c}$
gets enhanced when TM atoms with radius higher than the Fe atom are
substituted. It is reported that the application of external pressure
increases the $T_{c}$ in FeSe compounds \cite{Medvedev_11,Mizuguchi_16}.
If we correlate the increase in $T_{c}$ in our substituted compounds,
then we can assign a chemical pressure which will be equivalent to
an external pressure of $\sim0.5$\,GPa \cite{Mizuguchi_16}. If
we compare the maximum $T_{c}$ obtained for various TM metal substitutions
(Fig.~\ref{fig:Variation-of-}), we can bring in a correlation between
the enhancement of $T_{c}$ and the magnetic moment of the TM ion.
$T_{c}$ is found to increase as the moment decreases. The absence
of superconductivity in Co and Ni substituted compounds may be associated
with the dependence of ionic radius of the TM on the $T_{c}$. Since
they have ionic radius less than that of the Fe, it is possible that
they do not exert enough chemical pressure for the appearance of superconductivity.
Thus we can conclude that all the three effects, the amount of hexagonal
phase (chemical pressure), ionic radius and magnetic moment of the
substituted TM, may play a role in the enhancement/suppression of
$T_{c}$ and hence the superconducting properties. 

\section{Conclusions}

We have studied the effect of nominal substitution of the transition
metal ($T$ = Ti, V, Cr, Mn, Fe, Co and Ni) in place of excess Fe
in Fe$_{1+x}$Se superconductor. All the Fe$T_{x}$Se samples were
synthesized successfully via a single step solid state reaction method,
followed by quenching in LN2 from 360\,$^{\circ}$C. All the presented
transition metal (TM) substituted samples have tetragonal and hexagonal
phases. The superconducting transition temperature is enhanced by
10\% to 30\% when TMs with higher ionic radius compared to that of
the Fe is substituted. However, the substitution of the lower ionic
radius TM suppresses the $T_{c}$. The optimal concentration for the
highest $T_{c}$ is found to be different for different TM substitutions.
Both type of charge carriers were found to be present in the Fe- and
Cr-excess samples, however, other TM substituted samples show positive
sign of Seebeck coefficient throughout temperature range that indicates
holes as majority charge carriers. In brief, we can conclude that
the amount of hexagonal phase (chemical pressure), ionic radius as
well as the magnetic moment of the substituted TM may play a role
in the enhancement of $T_{c}$ and hence the superconducting properties
in the Fe-11 compound. 
\begin{acknowledgments}
CVT would like to acknowledge the Department of Science and Technology
for partial support through the project IR/S2/PU-10/2006. AKY would
like to thank CSIR, India for SRF grant. ADT acknowledges the Indian
Institute of Technology, Bombay for partial financial support during
part of this work and the Indian Institute of Technology, Patna for
seed grant.\end{acknowledgments}


\begin{thebibliography}{10}
\bibitem{Kamihara_1}Y. Kamihara, T. Watanabe, M. Hirano, and H. Hosono,
J. Am. Chem. Soc\emph{ }130 (2008) 3296.

\bibitem{Rottar_2} M. Rotter, M. Tegel, and D. Johrendt, Phys. Rev.
Lett. 101 (2008)107006.

\bibitem{Wang_3}X. C. Wang, Q. Q. Liu, Y. X. Lv, W. B. Gao,X. L.
Yang, R. C. Yu, F. Y. Li, and C. Q. Jin, Solid State Commun. 148 (2008)
538. 

\bibitem{Hsu_4} F. C. Hsu et al\emph{.} Proc. Natl Acad. Sci.\emph{
}105 (2008) 14262.

\bibitem{Ogino_5} H. Ogino, Y. Matsumara, Y. Katsura, K. Ushiyama,
S. Horii, K. Kishio, and J. I. Shimoyama, Supercond. Sci. Technol.
22 (2009) 075008.

\bibitem{Bao_6} W. Bao, Q. Huang, G. F. Chen, M. A. Green, D. M.
Wang, J. B. He, X. Q. Wang, and Y. Qiu, Chin. Phys. Lett. 28 (2011a)
086104.

\bibitem{Stewart_7}G. R. Stewart, Rev. Mod. Phys. 83 (2011) 1589.

\bibitem{Wang2_8}C. Wang \emph{et al.,} Europhys. Lett. 83 (2008
) 67006. 

\bibitem{Mizuguapl_9}Y. Mizuguchi \emph{et al.,} Appl. Phys. Lett.
93 (2008) 152505. 

\bibitem{Margadona_10}S. Margadonna, Y. Takabayashi, Y. Ohishi, Y.
Mizuguchi, Y. Takano, T. Kagayama, T. Nakagawa, M. Takata and K. Prassides,
Phys. Rev. B 80 (2009) 064506.

\bibitem{Medvedev_11} S. Medvedev \emph{et al}., Nature Mater. 8
( 2009) 630.

\bibitem{Fang_12} M. H. Fang, H. M. Pham, B. Qian, T. J. Liu, E.
K. Vehstedt, Y. Liu, L. Spinu, and Z. Q. Mao, \textit{\emph{Phys.
Rev. B}} 78 ( 2008) 224503. 

\bibitem{Yeh_13} K. W. Yeh,\emph{ }\textit{et al}\textit{\emph{.,
Europhys. Lett.}} 84 (2008) 37002.

\bibitem{Sales_14} B. C. Sales, A. S. Sefat, M. A. McGuire, R. Y.
Jin and D. Mandrus, \textit{\emph{Phys. Rev. B}} 79 ( 2009) 094521. 

\bibitem{Taen_15}T. Taen, Y. Tsuchiya, Y. Nakajima and T. Tamegai\textit{\emph{,
Phys. Rev. B}} 80 (2009) 092502. 

\bibitem{Mizuguchi_16}Y. Mizuguchi, F. Tomioka, S. Tsuda, T. Yamaguchi
and Y. Takano, J. Phys. Soc. Japan 78 (2009) 074712.

\bibitem{Wu_17}M. K. Wu, \emph{et al}., Physica C 469 (2009) 340.

\bibitem{guo18} J. Guo, S. Jin, G. Wang, S. Wang, K. Zhu, T. Zhou,
M. He, and X. Chen, Phys. Rev. B 82 (2010) 180520. 

\bibitem{Wang19} A. F. Wang\emph{ et al.,} Phys. Rev. B 83 (2011)
060512. 

\bibitem{Kazakov20}S. M. Kazakov, \emph{et al}., Chem. Mater. 23
(2011) 4311.

\bibitem{Anilprb_21}A. K. Yadav, A. D. Thakur and C. V. Tomy, \textit{\emph{Phys.
Rev. B}} 87 (2013) 174524.

\bibitem{Anilsolid_22}A. K. Yadav, A. D. Thakur, C. V. Tomy\textit{\emph{,
Solid State Commun.}} 151 (2011) 557.

\bibitem{Okamoto_23}H. Okamoto, J. Phase Equilib\emph{.} 12 (1991)
383.

\bibitem{McQueen_24} T. M. McQueen \emph{et al}., Phys. Rev. B 79
(2009) 014522.

\bibitem{Zhang_25}S. B. Zhang, H. C. Lei, X. D. Zhu, G. Li, B. S.
Wang, L. J. Li, X. B. Zhu, W. H. Song, Z. R. Yang, Y. P. Sun, \textit{\emph{Physica
C}} 469 (2009) 1958.

\bibitem{Zhang(22SST)_26}S. B. Zhang, \textit{et al.,} Supercond.
Sci. Technol. 22 (2009) 015020.

\bibitem{Zhang(T@H SST)_27} S. B. Zhang \emph{et al}., Supercond.
Sci. Technol.\emph{ }22 (2009) 075016.

\bibitem{Hein_28} R. H. Hein, Phys. Rev. B 33 (1986) 7539.

\bibitem{Song_29}Y. J. Song, J. B. Hong, B. H. Min, K. J. Lee, M.
H. Jung, J. S. Rhyee, Y. S. Kwon, J. Korean, Phys. Soc\emph{.} 59
(2011) 312.

\bibitem{Werthamer_30}N. R. Werthamer, E. Helfand and P. C. Hohenberg,
\textit{\emph{Phys. Rev.}} 147 (1966) 295.

\bibitem{Clogston_31}A. M. Clogston,\emph{ }\textit{\emph{Phys. Rev.
Lett}}\emph{.} 9 (1962) 266.

\bibitem{Chen} G. F. Chen, Z. G.Chen, J. Dong, W. Z. Hu, G. Li, X.
D. Zhang, P. Zheng, J. L. Luo and N. L. Wang, Phys. Rev. B 79 (2009)
140509. 

\bibitem{Khim} S. Khim, J. W. Kim, E. S. Choi, Y. Bang, M. Nohara,
H. Takagi, K. H. Kim, Phys. Rev. B. 81 (2010) 184511.

\bibitem{Lei2_32}H. C. Lei, R. W. Hu, E. S. Choi, J. B. Warren and
C. Petrovic, \textit{\emph{Phys. Rev. B}} 81 (2010) 184522.

\bibitem{Mizuguchi1pl_33}Y. Mizuguchi, F. Tomioka, S. Tsuda, T. Yamaguchi,
Y. Takano, \textit{\emph{Appl. Phys. Lett.}} 94 (2009) 012503.

\bibitem{Yang_34} X. Zhu, H. Yang, L. Fang, G. Mu, H. H. Wen\textit{\emph{,
Supercond. Sci. Technol.}} 21 (2008) 105001.

\bibitem{Wen_35}H. H. Wen, G. Mu, L. Fang, H. Yang, X. Y. Zhu, \textit{\emph{Europhys.
Lett.}}\emph{ }82 (2008) 17009.

\bibitem{Subedi}A. Subedi, L. Zhang, D. J. Singh, M. H. Du, Phys.
Rev. B\emph{ }78 (2008) 134514.

\bibitem{Johnston} D. C. Johnston, Advances in Physics 6 (2010) 803.

\bibitem{Xia}Y. Xia, D. Qian, L. Wray, D. Hsieh, G. F. Chen, J. L.
Luo, N. L. Wang, M. Z. Hasan, Phys. Rev. Lett.103 (2009) 037002.

\bibitem{Pomjakushina}E. Pomjakushina, K. Conder, V. Pomjakushin,
M. Bendele, R. Khasanov, Phys. Rev. B 80 (2009) 024517.

\bibitem{Braithwaite_36}D. Braithwaite, B. Salce, G. Lapertot, F.
Bourdarot, C. Marin, D. Aoki, M. Hanfland, J.\emph{ }\textit{\emph{Phys.:
Condens. Matter}}\textit{ }21 (2009) 232202.\end{thebibliography}
\end{document}